\documentclass[twocolumn,superscriptaddress,aps,longbibliography,showpacs,preprintnumbers,amsmath,amssymb,floatfix]{revtex4-1}
\usepackage{titlesec}
\usepackage{amsmath}
\usepackage{graphicx}
\usepackage[ansinew]{inputenc}
\usepackage{array}
\usepackage{color}
\usepackage{amsxtra}
\usepackage{amstext}
\usepackage{amssymb}
\usepackage{latexsym}
\usepackage{gensymb}
\usepackage{dsfont}
\usepackage{braket}
\usepackage[caption=false]{subfig}
\usepackage[makeroom]{cancel}

\usepackage{dcolumn}
\usepackage{bm}
\usepackage{bbm}
\usepackage{braket}
\usepackage{mathrsfs}
\usepackage{amstext}
\usepackage{euscript}
\usepackage{txfonts}

\usepackage[unicode=true,
 bookmarks=false,
 breaklinks=false,pdfborder={0 0 1},backref=false,colorlinks=true,citecolor=blue]
 {hyperref}

\makeatletter
\@ifundefined{textcolor}{}
{%
 \definecolor{BLACK}{gray}{0}
 \definecolor{WHITE}{gray}{1}
 \definecolor{RED}{rgb}{1,0,0}
 \definecolor{GREEN}{rgb}{0,1,0}
 \definecolor{BLUE}{rgb}{0,0,1}
 \definecolor{CYAN}{cmyk}{1,0,0,0}
 \definecolor{MAGENTA}{cmyk}{0,1,0,0}
 \definecolor{YELLOW}{cmyk}{0,0,1,0}
}

\makeatletter
\renewcommand*\env@matrix[1][*\c@MaxMatrixCols c]{%
  \hskip -\arraycolsep
  \let\@ifnextchar\new@ifnextchar
  \array{#1}}
\makeatother

\newcommand{\cref}[1]{Ref.\,\cite{#1}}

\makeatother

\begin{document}


\title{A stroboscopic approach to trapped-ion quantum information processing with squeezed phonons}
\date{\today}
\author{Wenchao Ge}
\affiliation{United States Army Research Laboratory, Adelphi, Maryland 20783, USA}
\affiliation{The Institute for Research in Electronics and Applied Physics (IREAP), College Park, Maryland 20740, USA}
\affiliation{Institute for Quantum Science and Engineering (IQSE) and Department of Physics and Astronomy, Texas A\&M University, College Station, TX 77843-4242, USA}

\author{Brian C. Sawyer}
\affiliation{Georgia Tech Research Institute, Atlanta, Georgia 30332, USA}

\author{Joseph W. Britton}
\affiliation{United States Army Research Laboratory, Adelphi, Maryland 20783, USA}

\author{Kurt Jacobs}
\affiliation{United States Army Research Laboratory, Adelphi, Maryland 20783, USA}
\affiliation{Department of Physics, University of Massachusetts at Boston, Boston, Massachusetts 02125, USA}
\affiliation{Hearne Institute for Theoretical Physics, Louisiana State University, Baton Rouge, Louisiana 70803, USA}

\author{Michael Foss-Feig}
\affiliation{United States Army Research Laboratory, Adelphi, Maryland 20783, USA}
\affiliation{Joint Quantum Institute, NIST/University of Maryland, College Park, Maryland 20742, USA}
\affiliation{Joint Center for Quantum Information and Computer Science, NIST/University of Maryland, College Park, Maryland 20742, USA}

\author{John J. Bollinger}
\affiliation{National Institute of Standards and Technology, Boulder, Colorado 80305, USA,}

\begin{abstract}
In trapped-ion quantum information processing, interactions between spins (qubits) are mediated by collective modes of motion of an ion crystal.  While there are many different experimental strategies to design such interactions, they all face both technical and fundamental limitations to the achievable coherent interaction strength.  In general, obtaining strong interactions and fast gates is an ongoing challenge.  Here, we extend previous work [Phys. Rev. Lett. \textbf{112}, 030501 (2019)] and present a general strategy for enhancing the interaction strengths in trapped-ion systems via parametric amplification of the ions' motion. Specifically, we propose a stroboscopic protocol using alternating applications of parametric amplification and spin-motion coupling. In comparison with the previous work, we show that the current protocol can lead to larger enhancements in the coherent interaction that increase exponentially with the gate time. \end{abstract}

\maketitle

\section{Introduction}
Trapped ions are a well-established platform for numerous quantum applications, including quantum computation \cite{PhysRevLett.74.4091}, quantum simulation \cite{Blatt:2012aa}, and quantum metrology \cite{PhysRevA.46.R6797}.  One key advantage of trapped ions compared to other quantum platforms is the relatively large ratio between achievable coherent interaction rates and decoherence rates.  This large ratio enables high-fidelity single qubit and two-qubit operations \cite{Ballance:2016, Gaebler:2016aa}, and renders ion traps amongst the most capable platforms for generating large amounts of useful entanglement  \cite{Bohnet:2016aa, Zhang_2017}.

Nevertheless, trapped ions are not without their own specific challenges and shortcomings \cite{PhysRevA.75.042329,PhysRevLett.105.200401,PhysRevA.97.052301}. In absolute terms, interaction strengths are small compared to other technologies based on solid-state or superconducting qubits, leading to relatively slow gates.  And while the achievable gate fidelities are state-of-the-art, significant improvements are desirable and most likely necessary for scalable quantum computation.  In all trapped ion experiments to date, spin-spin interactions are generated by controllably coupling internal states of the ions to collective motional modes of a crystal in which they sit through the application of a spin-dependent force \cite{PhysRevLett.75.4714, Ospelkaus:2011aa, maggrad2}.  Regardless of how the spin-dependent force is created (either through optical dipole forces or magnetic field gradients), there are both technical and fundamental limitations to how strong it can be---and therefore how fast gates can be accomplished---without degrading the gate fidelity.  For example, the spin-dependent force is often limited technically by the availability of the laser power (in optical gates) or the current that can be driven into a thin trap electrode (in microwave gates).  Other more fundamental tensions between gate speed and fidelity exist, e.g. arising from off-resonant coupling to auxiliary motional modes \cite{PhysRevLett.112.190502,PhysRevLett.114.120502, PhysRevLett.120.020501}.

\begin{figure}[t]
\leavevmode\includegraphics[width = 1 \columnwidth]{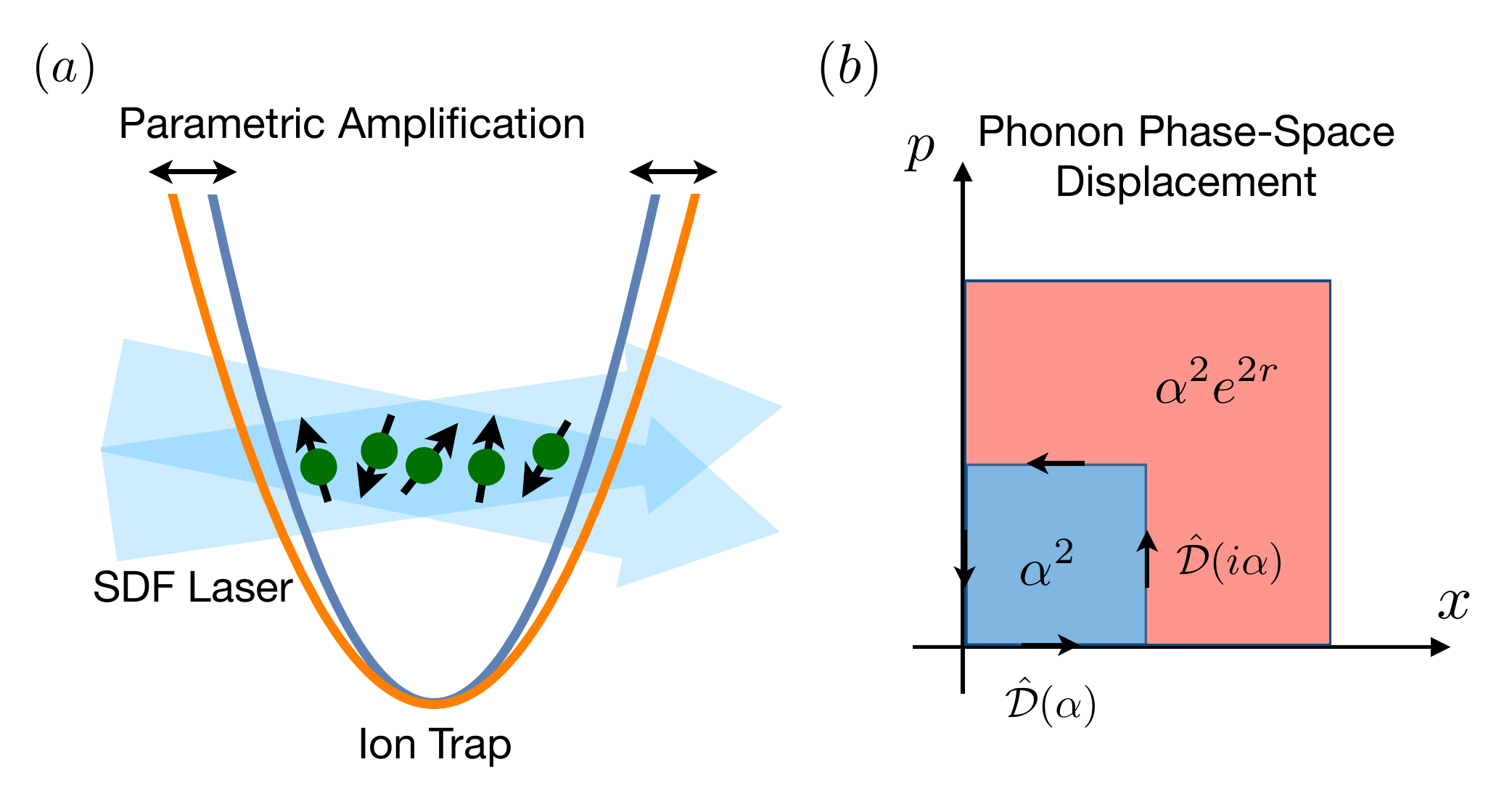}
\caption{(a) Trapped ion setup with a spin-dependent optical dipole force from off-resonant laser beams and parametric amplification from a modulation of the trap potential. (b) A square protocol using a SDF with (red square) and without (blue square) parametric amplification.} 
\label{fig:scheme} 
\end{figure}

Recently we proposed a new mechanism \cite{GePRL2019} to use parametric amplification (PA), via modulation of the ions' trapping potential at twice the target motional mode frequency \cite{Heinzen1990}, to enhance the spin-spin interaction strength at a fixed strength of the spin dependent force. The basic idea of using a parametric drive can be summarized as follows. The effective spin-spin interaction arises from the spin-dependent acquisition of geometric phase accrued through displacements of the mechanical modes of the ion crystal \cite{Sorensen00, Leibfried03, G-Ripoll05}. These spin-dependent displacements are seeded by a spin-dependent force (SDF), and further amplified spin-independently by a parametric drive that modulates the trapping potential (see Fig.  \ref{fig:scheme}). As a result of this amplification, the ion can acquire an enhanced spin-dependent geometric phase per unit of time. In Ref. \cite{GePRL2019}, we proposed a continuous protocol in which a spin-dependent force and the parametric drive were applied at the same time, and showed that the total Hamiltonian can be mapped to the original spin-dependent Hamiltonian with a spin-motion coupling strength that grows algebraically in the strength of the parametric drive. Here, we extend this idea by studying a protocol where the PA and the SDF are applied in non-overlapping pulses. Our results show that a stroboscopic protocol of this sort can have an exponential enhancement in the geometric phase, leading to significantly greater enhancements at long interaction times. This approach may be useful in other related systems with boson-mediated interactions modified by PA, such as phonon-mediated superconductivity \cite{Babadi2017}, optomechanics \cite{Lemonde:2016aa} and cavity or circuit QED  \cite{PhysRevX.7.021041, arenz2018hamiltonian}.

The paper is organized as follows. In Sec. II, we first introduce the basics of quantum simulations with trapped ions and discuss a stroboscopic protocol using only a SDF. We then summarize some of the challenges in generating coherent spin-spin interactions
with spin-dependent forces and trapped ions. In Sec. III, we present the stroboscopic protocol with parametric amplification, including discussions on the process of parametric amplification, the details of a ``square" stroboscopic protocol with both PA and SDF, and an error analysis. We summarize our work and provide an outlook that includes a discussion of some of the trade-offs in implementing PA in Sec. IV. For better readability, we defer detailed derivations and further explanations to the appendix.

\section{Overview of quantum simulation with trapped ions}

\subsection{The effective Hamiltonian}
The Hamiltonian describing a crystal of $N$ trapped ions with two long-lived internal states can be written as ($\hbar=1$) \cite{Porras2004}
\begin{align}
\hat{\mathcal{H}}_{\rm ions}=\sum_{m}\omega_{m}\hat{a}^{\dagger}_{m}\hat{a}^{\phantom\dagger}_{m}+\frac{\omega_z}{2}\sum_{i=1}^{N}\hat{\sigma}^{(i)}_z,
\end{align}
where $\hat{a}_{m}$ is the annihilation operator of the mechanical mode with frequency $\omega_m$, and $\hat{\sigma}^{(i)}_z$ is $z$-Pauli matrix for the $i$th spin with the qubit frequency splitting $\omega_{z}$. 

The most common way to generate entanglement between trapped-ion spins is to create effective spin-spin interactions mediated by the collective ion motion through the application of an oscillating SDF. This is most often achieved using noncopropagating lasers to either drive stimulated Raman transitions \cite{Molmer:1999aa} or to generate an AC stark shift on the qubit transition that is spatially varying \cite{Milburn:2000aa}. Note that recently it has also been achieved using strong magnetic field gradients in surface-electrode traps \cite{Ospelkaus:2011aa,maggrad2}. Without loss of generality, we consider a SDF that couples the internal spin $\hat{\sigma}^{(i)}_z$ to the $z$-direction of the external ion motion. In the Lamb-Dicke regime and in the rotating frame of the qubit and the normal modes, the SDF Hamiltonian \cite{Britton2012} is 
\begin{eqnarray}
\hat{\mathcal{H}}_{\text{S}}(t)= F\sin\left(\mu t\right)\sum_{i=1}^N\hat{z}_i(t)\hat{\sigma}^{(i)}_z.\label{eq:NHodf}
\end{eqnarray}
The ion position operators can be written $\hat{z}_i(t)=\sum_{m=1}^Nb_{i,m}z_{0m}\big(e^{-i\omega_m t}\hat{a}_m+e^{i\omega_m t}\hat{a}_m^{\dagger}\big)$, where $z_{0m}=\sqrt{\hbar/2M\omega_m}$, $M$ is the ion mass, and $b_{i,m}$ are the normal-mode transformation matrix elements obeying $\sum_{i=1}^Nb_{i,m}b_{i,l}=\delta_{ml}$ and $\sum_{m=1}^Nb_{i,m}b_{j,m}=\delta_{ij}$ \cite{James:1998aa}. When the SDF is generated by two noncopropagating lasers with wave vector difference $\Delta k$, the Lamb-Dicke regime requires $\Delta k \sqrt{\braket{\hat{z}^2_i(t)}}\ll 1$. 

The formal expression for the unitary evolution due to the Hamiltonian $\hat{\mathcal{H}}_{S}(t)$ is given by 
\begin{align}
\label{eq:unitary}
&\hat{\mathcal{U}}_{S}(t,t_0)=\mathcal{T}\exp\left(-i\int_{t_0}^t dt^{\prime}\hat{\mathcal{H}}_{S}(t^{\prime})\right)\nonumber\\
&= \exp\left(-i\int_{t_0}^t dt^{\prime}\hat{\mathcal{H}}_{S}(t^{\prime})-\frac{1}{2}\int_{t_0}^t dt^{\prime}\int_{t_0}^{t^{\prime}} dt^{\prime\prime}\left[\hat{\mathcal{H}}_{S}(t^{\prime}),\hat{\mathcal{H}}_{S}(t^{\prime\prime})\right]+\cdots\right)\nonumber\\
&=\exp\left[\sum_{i=1}^N\hat{\phi}_i(t,t_0)\hat{\sigma}_{z}^{(i)}\right]\exp\left[i\sum_{i,j=1}^N\Phi_{ij}(t,t_0)\hat{\sigma}_{z}^{(i)}\hat{\sigma}_{z}^{(j)}\right]
\end{align}
where $\mathcal{T}$ is the time-ordering operator. The second line of the above equation is obtained using Magnus' expansion. Because $\left[\hat{\mathcal{H}}_{S}(t_1),\left[\hat{\mathcal{H}}_{S}(t_2),\hat{\mathcal{H}}_{S}(t_3)\right]\right]=0$ \cite{Zhu:2003aa}, we arrive at the third line, where the first term represents spin-motion coupling and the second term is the effective spin-spin interaction mediated by the mechanical motion. The spin-motion coupling results in a spin-dependent displacement of the mechanical motion with $\hat{\phi}_i(t,t_0)=\sum_{m=1}^N\left[\alpha_{i,m}(t,t_0)\hat{a}^{\dagger}_m-\alpha_{i,m}^{\ast}(t,t_0)\hat{a}_m\right]$ such that the displacement of the $m$th motional mode from the $i$th ion follows the trajectory $\pm\alpha_{i,m}(t,t_0)$ for $\ket{\pm}_i$, respectively. Here $\sigma^{(i)}_z \ket{\pm}_i = \pm\ket{\pm}_i$ and  
\begin{align}
\alpha_{i,m}(t,t_0)&=-i2b_{i,m}f_m\int_{t_0}^{t}\sin(\mu t^{\prime})e^{i\omega_m t^{\prime}}dt^{\prime},\label{eq:fma}
\end{align}
with $f_m\equiv Fz_{0m}/2$. If the counter rotating term is neglected, then at a time given by an integer multiple of $2\pi/\left|\mu-\omega_m\right|$, the displacement of the $m$th mode $\alpha_{im}= 0$, i.e., the spin and motion are disentangled.  The phase $\Phi_{ij}$ is given by
\begin{align}
\Phi_{ij}(t,t_0)&=\text{Im}\left[\sum_m \int_{t_0}^{t}\frac{d\alpha_{j,m}(t^{\prime},t_0)}{dt^{\prime}}\alpha^{\ast}_{i,m}(t^{\prime},t_0)dt^{\prime} \right].
\label{eq:ss}
\end{align}
It is called a geometric phase because it is twice the sum of the geometric area accumulated by the trajectory of each mode. As a simple example of how the unitary dynamics describing a spin-spin interaction can be engineered, we consider a stroboscopic protocol that results in a square trajectory (as indicated by the arrows in Fig. \ref{fig:scheme} (b)) with a SDF that is resonant with a single mode, e.g., the center-of-mass motion corresponding to $m=1$ and $\mu=\omega_1$. This can be achieved by changing the phase of the SDF by $\pi/2$ after every $T/4$, where $T$ is the total evolution time. In place of Eq. \eqref{eq:unitary}, the unitary evolution of each segment can be now more simply represented by $\hat{\mathcal{D}}\left(\sum_{i=1}^N\alpha_{i,1}\hat{\sigma}_z^{(i)}\right)$ with $\alpha_{i,1}=f_1T/\left(4\sqrt{N}\right)$ and  the displacement operator $\hat{\mathcal{D}}(\alpha)\equiv\exp\left(\alpha \hat{a}^{\dagger}-\alpha^{\ast}\hat{a}\right)$. The entire protocol is given by the following evolution
\begin{align}
&\hat{\mathcal{D}}\left(-i\sum_{i=1}^N\alpha_{i,1}\hat{\sigma}_z^{(i)}\right)\hat{\mathcal{D}}\left(-\sum_{i=1}^N\alpha_{i,1}\hat{\sigma}_z^{(i)}\right)\hat{\mathcal{D}}\left(i\sum_{i=1}^N\alpha_{i,1}\hat{\sigma}_z^{(i)}\right)\hat{\mathcal{D}}\left(\sum_{i=1}^N\alpha_{i,1}\hat{\sigma}_z^{(i)}\right)\nonumber\\
&=\exp\left[i\Phi\sum_{i,j=1}^N\hat{\sigma}_{z}^{(i)}\hat{\sigma}_{z}^{(j)}\right],
\label{eq:sp-sdf}
\end{align}
where $\Phi=2\left(f_1T/4\right)^2/N$ and we have used $\hat{\mathcal{D}}(\alpha)\hat{\mathcal{D}}(\beta)=\hat{\mathcal{D}}(\alpha+\beta)\exp\left(i\text{Im}[\alpha\beta^{\ast}]\right)$ \cite{SZ}. 

\subsection{Challenges in generating coherent spin-spin interactions with spin-dependent forces}
\subsubsection{Gate time}
As can be seen from Eq. \eqref{eq:sp-sdf}, the minimum gate time for a particular entangled spin state obtained with a geometric phase $\Phi$  is limited by the achievable size of the spin dependent force $f$ as
\begin{align}
T\propto \frac{\sqrt{\Phi}}{f}.
\label{eq:gatetime}
\end{align}
Here, we drop the subscript in $f_1$ for simplicity. This relation can be a technical limitation because larger laser power may not be available or it may not be possible to drive a larger current through a thin electrode of a trap. 

\subsubsection{Spontaneous photon scattering}
Even assuming we have access to stronger laser powers, this may not help to reduce the decoherence. For example, in experiments employing optical dipole forces to generate the SDF,  the spontaneous photon scattering rate is proportional to the laser power of the driving beams. In some experiments, the dominant source of decoherence is due to photon scattering and occurs at a rate $\Gamma\propto f$ \cite{PhysRevA.75.042329,PhysRevLett.105.200401}. Therefore, preparation of a particular entangled spin state is accompanied with the accumulated decoherence  that is independent of $f$
\begin{align}
\Gamma T\propto \sqrt{\Phi}.
\end{align}

\subsubsection{Residual spin-motion entanglement}
The spin-spin interactions may also suffer decoherence from residual motional displacements that produce spin-motion entanglement. This happens when the motional modes do not close a loop in phase-space at the gate time, i.e. $\alpha_{im}(T)\ne 0$. The origin of the residual displacements can be categorized into two cases: 1) due to imperfect control of system parameters, such as time and frequency; 2) due to coupling to multiple modes with different frequencies, such that not all of them return to the origin in phase space at the same time. This problem becomes more severe when the number of ions increases.  Parametric amplification can mitigate this second source of residual spin-motion entanglement.  

\section{Enhancing coherent interaction with parametric amplification}
In Ref. \cite{GePRL2019} we proposed to use a spin-independent modulation of the trapping potential in order to enhance the coherent spin-spin interaction and mitigate the limitations discussed above. The spin-dependent displacements of the phonons are seeded by the spin-dependent force, and further amplified by a simultaneously applied spin-independent parametric drive. In an extension to this previous work \cite{GePRL2019}, we now consider alternating the application of a spin-dependent force with interspersed periods of trap modulation. In this way, the mechanical modes can once again acquire an enhanced geometric phase per unit of time (the red square in Fig. \ref{fig:scheme} (b)). For the square protocol, we quantify the enhancement by the factor $G$ defined through the relation
\begin{align}
\Phi= G\frac{2}{N}\left(\frac{fT}{4}\right)^2.\label{eq:enhanced}
\end{align}

\subsection{Parametric amplification}

Before presenting the protocol, we specify the form of PA discussed here. The PA process amplifies motion along one quadrature of a harmonic oscillator and attenuates motion along the conjugate quadrature. In trapped ions, this can be done by driving the appropriate ion-trap electrodes at close to twice the motional resonance frequency \cite{Heinzen1990,Burd:2019aa}.  As shown in the appendix, a parametric drive on all the ions can be transformed into a summation of parametric amplification on each mode separately.  Applying the rotating-wave approximation (RWA), we obtain
\begin{eqnarray}
\hat{\mathcal{H}}_{PA}(t)=\sum_{m=1}^N\hat{\mathcal{H}}_{PA}^{(m)}(t), 
\label{eq:PA}
\end{eqnarray} 
with 
\begin{eqnarray}
\hat{\mathcal{H}}_{PA}^{(m)}(t)=-\frac{i}{2}\left(g_me^{-i2\Delta_mt}a_m^{\dagger2}-g_m^{\ast}e^{i2\Delta_mt}a_m^2\right)
\end{eqnarray} 
where $g_m\equiv \frac{e|V|e^{i\theta}}{M\omega_md^2_T}$, $\Delta_m=(\omega_p/2-\omega_m)$, and $\omega_p$ is the parametric modulation frequency. Here $V$ is the voltage of the parametric drive and $d_T$ is a characteristic trap dimension as discussed in Appendix A. Since the Hamiltonian is time-dependent, the unitary evolution is given by
\begin{eqnarray}
\hat{\mathcal{U}}_{P}(t,t_0)&=&\mathcal{T}\exp\left(-i\int_{t_0}^t dt^{\prime}\hat{\mathcal{H}}_{PA}(t^{\prime})\right)\nonumber\\
&=&\prod_{m=1}^N\mathcal{T}\exp\left(-i\int_{t_0}^t dt^{\prime}\hat{\mathcal{H}}_{PA}^{(m)}(t^{\prime})\right).
\end{eqnarray}
Therefore, the PA unitary is just a product of unitaries for each mode that can be acted separately on the system. In particular, we consider the situation where PA is resonant with a single target mode, e.g., the center-of-mass motion, such that the parametric amplification is more important on the target mode than the other "spectator" modes. Conditions for the validity of this single mode analysis are discussed in Section III.C.1.

\begin{figure}[t]
\leavevmode\includegraphics[width = .95 \columnwidth]{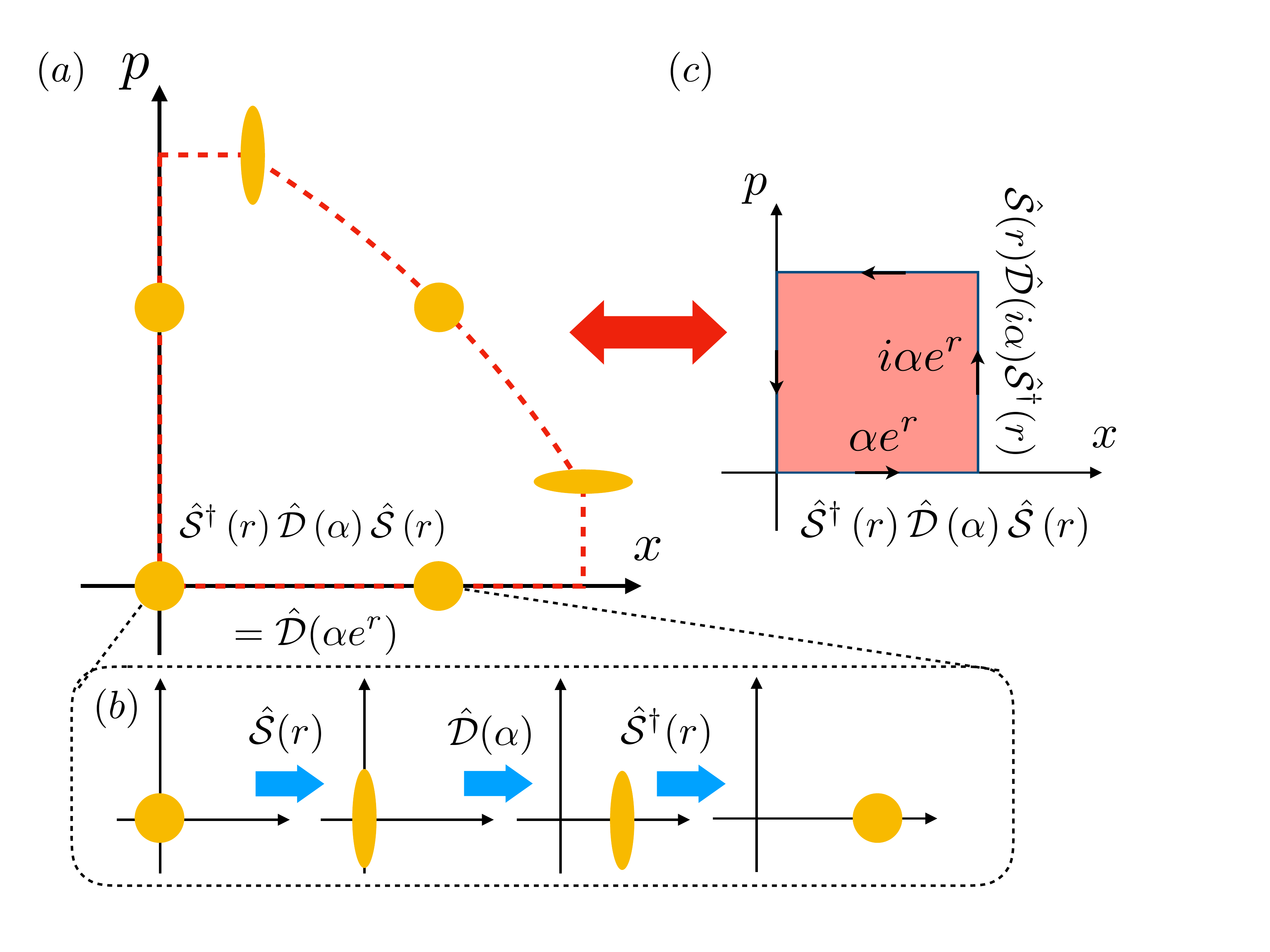}
\caption{(a) Actual phase-space trajectory due to the SDF and PA. (b) State evolution under the operations $\hat{\mathcal{S}}(r)$, $\hat{\mathcal{D}}(\alpha)$, and $\hat{\mathcal{S}}^{\dagger}(r)$ with circular discs representing coherent states and elliptical discs representing squeezed coherent states. (c) Effective phase-space displacements for the enhanced geometric area.} 
\label{fig:sp} 
\end{figure}

In this case $\omega_p=2\mu=2\omega_1$ and the PA unitary can be described by a squeezing operation $\hat{\mathcal{S}}\left(\xi\right)=\exp\left(\frac{1}{2}\xi^{\ast} \hat{a}^2-\frac{1}{2}\xi \hat{a}^{\dagger2}\right)$ with $\xi=g(t-t_0)\equiv re^{i\theta}$, where we take $g_1\equiv g$. The parametric driving squeezes the quadrature $(\hat{a}e^{-i\theta/2}+\hat{a}^{\dagger}e^{i\theta/2})/\sqrt{2}$ and amplifies the other quadrature $-i(\hat{a}e^{-i\theta/2}-\hat{a}^{\dagger}e^{i\theta/2})/\sqrt{2}$, which are rotated by $\theta/2$ compared to the usual quadratures. In contrast to spin-dependent forces, which are linear in $\hat{a}$ and $\hat{a}^\dagger$,  the squeezing operation is quadratic in $\hat{a}$ and $\hat{a}^{\dagger}$, and consequently the phase acquired by a spin state traversing the parametrically driven path in Fig. \ref{fig:sp} (a) can no longer be calculated from the area enclosed by this path. Instead, one has to convert the squeezing into effective amplified displacements before calculating the geometric area (see Fig. \ref{fig:sp} and discussion in the next Section). This gives certain restrictions on how to perform PA together with a SDF in order to restore the initial mechanical state with an enhanced geometric phase. 

\begin{figure}[t]
\leavevmode\includegraphics[width = .8 \columnwidth]{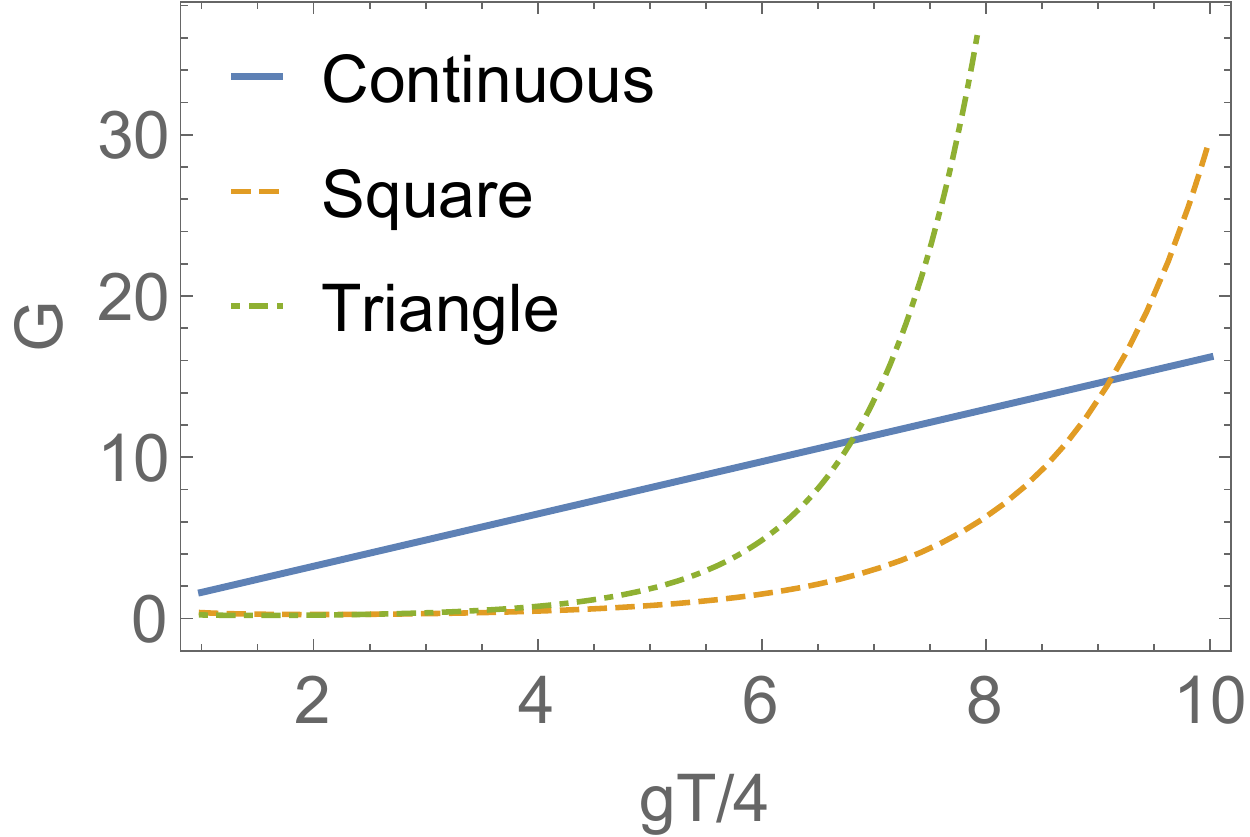}
\caption{Enhancement of geometric phase using a square protocol, a triangle protocol, and the continuous protocol in Ref. \cite{GePRL2019} for the same spin-dependent force $f$ and gate time $T$. In all three plots, we compare these protocols with the square protocol without the PA (Fig. \ref{fig:scheme} (b)).} 
\label{fig:enhancement} 
\end{figure} 

\subsection{Stroboscopic Protocol}
The degrees of freedom of a general stroboscopic protocol can be very large (see appendix), making it very difficult to find the optimal protocol for amplifying the geometric phase in a general situation. Here we focus on a simple but useful protocol that clearly illustrates how PA can enhance the effective spin-spin interaction strength. Specifically, we consider a square trajectory for the SDF and insert a squeezing operation before and after each displacement. The interaction time of one displacement operation is $t_1$, the same for all four sides. The squeezing time $t_2$ is the same for each operation. The first displacement becomes
\begin{align}
\hat{\mathcal{S}}^{\dagger}(r)\hat{\mathcal{D}}(\alpha)\hat{\mathcal{S}}(r),\label{eq:disprotocol}
\end{align}
where $\alpha=ft_1/\sqrt{N}$ and $r=gt_2$. Note that we have omitted the dependence on the spin states in the displacement operation for simplicity. For an initial ground state of the center-of-mass motion $\ket{0}_{\text c}$, the evolution of the state due to the three operations in Eq. \eqref{eq:disprotocol} is plotted in Fig. \ref{fig:sp} (b): $\ket{0}_{\text c}\rightarrow \ket{r}_{\text c}\rightarrow \ket{\alpha, r}_{\text c}\rightarrow \ket{\alpha e^r}_{\text c}$. Here, $\ket{r}_{\text c}$ is a squeezed vacuum state, $\ket{\alpha, r}_{\text c}$ is a squeezed coherent state, and $\ket{\alpha e^r}_{\text c}$ is a coherent state with amplitude $\alpha e^r$. To get the final state, we used the relation $\hat{\mathcal{S}}^{\dagger}(r)\hat{\mathcal{D}}(\alpha)\hat{\mathcal{S}}(r)=\hat{\mathcal{D}}(\alpha e^r)$ \cite{SZ} for real values $\alpha$ and $r$. Therefore, a small displacement $\alpha$ is amplified to $\alpha e^r$. Recently, this amplification protocol has been used to improve sensing of a small displacement \cite{Frowis2016, Burd:2019aa}.

To enhance the second displacement, we apply $\hat{\mathcal{S}}(r)\hat{\mathcal{D}}(i\alpha)\hat{\mathcal{S}}^{\dagger}(r)=\hat{\mathcal{D}}(i\alpha e^r)$. Similarly, the remaining operations are given by $\hat{\mathcal{S}}(r)\hat{\mathcal{D}}(-i\alpha)\hat{\mathcal{S}}^{\dagger}(r)\hat{\mathcal{S}}^{\dagger}(r)\hat{\mathcal{D}}(-\alpha)\hat{\mathcal{S}}(r)=\hat{\mathcal{D}}(-i\alpha e^r)\hat{\mathcal{D}}(-\alpha e^r)$. The actual trajectory is represented by the dashed curve in Fig. \ref{fig:sp} (a), where we have also denoted the wave functions at several points. In particular, the four discs in Fig. \ref{fig:sp} (a) represent the states at the corners in the effective displacements  in Fig. \ref{fig:sp} (c). The completed protocol is described by
\begin{align}
\hat{\mathcal{D}}(-i\alpha e^r)\hat{\mathcal{D}}(-\alpha e^r)\hat{\mathcal{D}}(i\alpha e^r)\hat{\mathcal{D}}(\alpha e^r)=\exp\left(i2\alpha^2e^{2r}\right).\label{eq:squared}
\end{align}
We see that the motional state returns to its initial state despite its quadratures being amplified and squeezed along the trajectory, tracing out an effective geometric area of $\alpha^2e^{2r}$. For a fixed gate time $T=4t_1+8t_2$, we find the optimal geometric phase to be
\begin{align}
\Phi=\frac{2}{N}\left(\frac{fT}{4}\right)^2\frac{e^{gT/4-2}}{\left(gT/4\right)^2},
\label{eq:enhancement}
\end{align}
at the optimal time $t_1=2/g$. The geometric phase is enhanced exponentially by the factor $G=e^{gT/4-2}/\left(gT/4\right)^2$. This is the main result of our work. Specifically, with stroboscopic parametric amplification one can achieve an exponential enhancement in the geometric phase.

More generally, we consider the situation of amplifying a regular $n$-sided polygon by inserting a squeezing operation before and after each displacement. We derive the optimal enhanced geometric area to be
\begin{align}
\Phi(n)=\frac{2}{N}\left(\frac{fT}{4}\right)^2\frac{n\cot\left(\frac{\pi}{n}\right)e^{gT/n-2}}{4\left(gT/4\right)^2}.
\end{align}
We see that the factor $n\cot\left(\frac{\pi}{n}\right)$ increases polynomially as $n$ increases, while the exponential factor $e^{gT/n-2}$ drops exponentially. Therefore, it is useful to consider the square protocol $n=4$ and a triangle protocol for $n=3$. We plot the enhancement factor $G$ for both these protocols in Fig. \ref{fig:enhancement}. The triangle protocol gives a larger enhancement for a fixed gate time T. As a comparison, we also plot the enhancement factor using the continuous protocol of Ref. \cite{GePRL2019} for the same spin-dependent force $f$ and the same gate time $T$. We see that for $gT\lesssim 30$, the continuous protocol can give a better enhancement, while the stroboscopic protocol may be more advantageous for $gT\gtrsim 30$. Note that $gT$ is bounded by a modified Lamb-Dicke limit for our model to be valid, and therefore the actual upper value of $gT$ depends on the specific experimental setup (see Appendix D).

\begin{figure}[t]
\leavevmode\includegraphics[width = .8 \columnwidth]{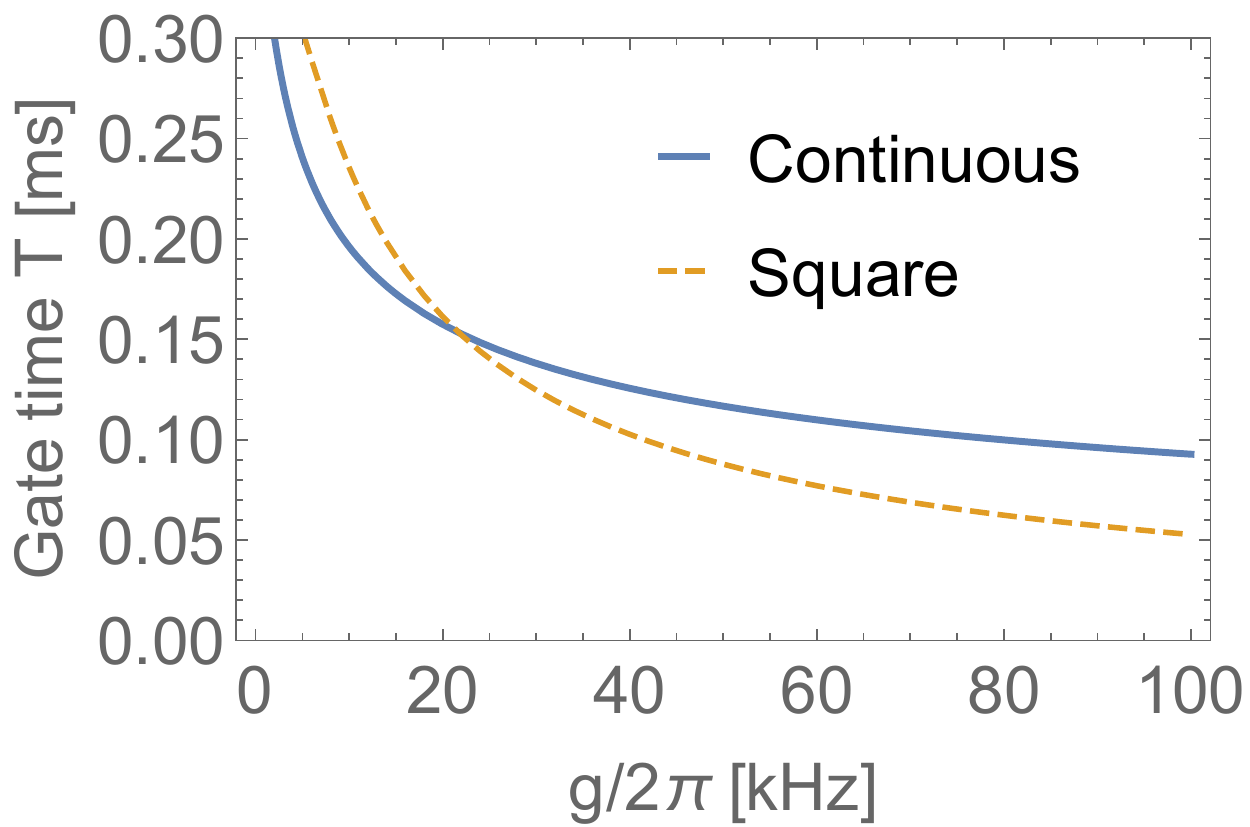}
\caption{Gate time reduction using a square protocol, and the continuous protocol in Ref. \cite{GePRL2019} for the same spin-dependent force $f$ and fixed geometric phase $\Phi$. The gate time without the PA is taken to be $0.4$ ms.} 
\label{fig:gt} 
\end{figure}

Above, we held fixed the SDF strength $f$ and the total interaction time and maximized the acquired geometric phase, which is equivalent to maximizing the spin-spin interaction strength at fixed $f$.  Alternatively, we could minimize the time required to achieve a particular geometric phase at fixed $f$. As we can see from Eq. \eqref{eq:enhancement}, the gate time can be reduced due to the exponential enhancement factor. For example, we plot the gate time as a function of $g$ for both the continuous protocol and the square protocol in Fig. \ref{fig:gt}.



\subsection{Error Analysis}
In our previous analysis, we assumed a single-mode approximation and perfect control of the experimental parameters. Here we analyze the leading sources of errors due to off-resonant coupling to unwanted modes, mode frequency fluctuations, relative phase fluctuations between the PA and SDF, and interaction time control fluctuations. 

\subsubsection{Off-resonant modes}
\begin{figure}[t]
\leavevmode\includegraphics[width = .8 \columnwidth]{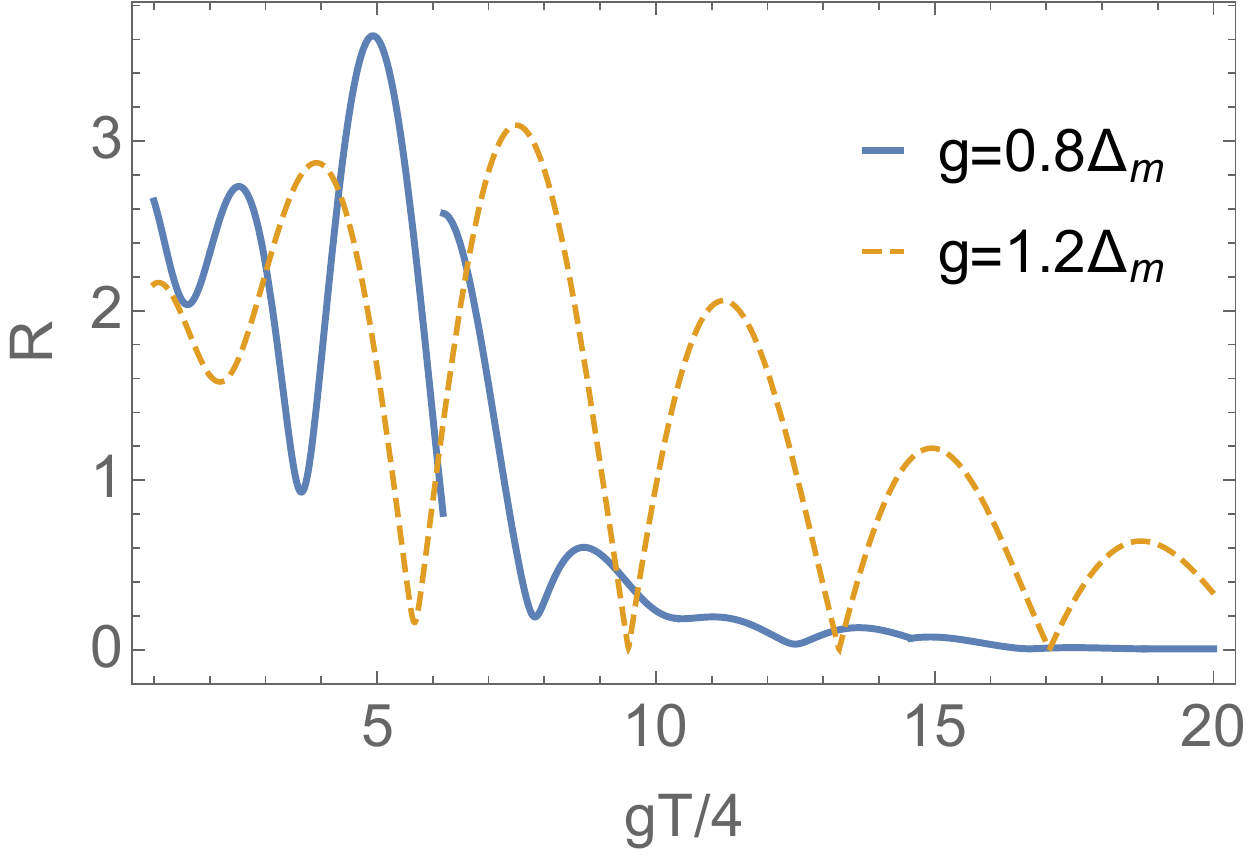}
\caption{The ratio of residual displacements with and without PA. The discontinuity in the solid curve is due to the jump in the squeezing angle that occurs for off-resonant squeezing as the squeezing time is increased. } 
\label{fig:ratio} 
\end{figure}

Off-resonant coupling to unwanted modes can lead to spin-motion entanglement if there is a residual displacement at the gate time. The residual displacement of unwanted modes results from both the off-resonant SDF and off-resonant PA. The value of the residual displacement can be numerically calculated (see Appendix C). Here we just give an estimate of the approximate size of the displacement by considering the operations in Eq. \eqref{eq:disprotocol} for the off-resonant modes. 
We define $R$ as the ratio of the residual displacements with and without using PA. We plot this ratio in Fig. \ref{fig:ratio} as a function of $gT/4$ for different values of the frequency offset $\Delta_m$ of the unwanted mode from the target mode. We observe that for $g\sim\Delta_m$, the residual displacements can be suppressed for $gT\gg1$ but the target mode is amplified. This condition gives an estimated bound on how strong the PA can be driven before amplifying the unwanted modes. We note that in principle these residual displacements may be eliminated with a more sophisticated protocol, i.e. with pulse shaping.

\subsubsection{Technical errors}
Anticipated sources of technical errors include mode frequency fluctuations, phase fluctuations between the SDF and PA, and imperfect timing control. We assume that the fluctuations are constant for the duration of a single experiment, but vary from one realization of the experiment to the next. Our analysis shows that the stroboscopic protocol is particularly sensitive to mode frequency fluctuations and phase fluctuations between the SDF and PA. Below, we quote the results and leave the detailed analysis for the appendix. 

\begin{figure}[t]
\leavevmode\includegraphics[width = .8 \columnwidth]{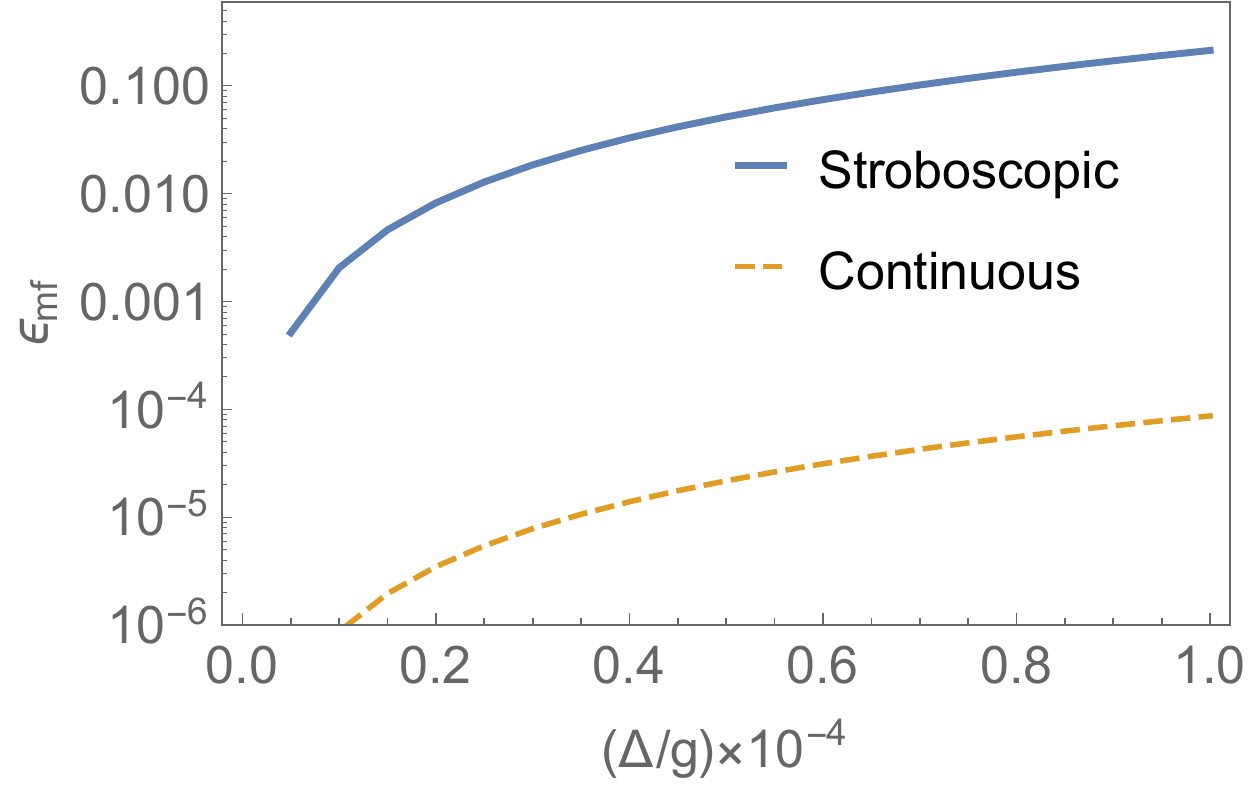}
\caption{The error due to mode frequency fluctuation $\Delta$ for both the continuous protocol and the stroboscopic protocol at $G=10$.} 
\label{fig:modeff} 
\end{figure}

Mode frequency fluctuations of size $\Delta$ lead to a rotation of the squeezing angle, which can result in an amplified residual displacement of (Appendix C)
\begin{align}
\Delta\alpha\approx C \left(\Delta/g\right)e^{2r}=C \left(\Delta/g\right)G(2r+2)^2,
\label{eq:ard}
\end{align}
where $r=gt_2$ and $C\approx 6$ determined from the numerical simulation in Fig. \ref{fig:modeff}. The error in the gate fidelity is $\epsilon_{\text{mf}}=\left(\Delta \alpha\right)^2$. Phase fluctuations $\Delta \theta$ between the PA and the SDF can lead to the same order of error. As analyzed in the Appendix C, phase fluctuations can give rise to phase rotation on two neighboring PA operations, i.e., a phase mismatch between squeezing and antisqueezing, which can result in a similar residual displacement as in Eq. \eqref{eq:ard}.

Timing fluctuations in the action of the SDF result in $\Delta\alpha\propto \epsilon\sqrt{\Phi}$ and $\Delta\Phi= \epsilon\Phi$, where $\epsilon$ is a fractional error in the time interval for the application of the SDF. Timing fluctuations due to the action of the PA only lead to residual displacement, $\Delta\alpha\propto r \epsilon\sqrt{\Phi}$, while preserving the geometric phase to the first order.



\subsection{Comparison to the continuous protocol}
In our recent work \cite{GePRL2019}, we presented a continuous protocol where the SDF and PA are applied simultaneously. A nice feature of the continuous protocol is that we obtain a total Hamiltonian that can be mapped to the original SDF-only Hamiltonian with an enhanced coupling strength. This can be useful for studies on spin-motion coupling \cite{Lewis-Swan:2019aa}. Additionally, the gate time is not limited by the motional mode splitting and the enhancement is more advantageous for $gT\lesssim 30$ as shown in Fig. \ref{fig:enhancement}. Like the stroboscopic protocol, the continuous protocol is also sensitive to mode frequency fluctuations. For comparison, we quote the error in the gate fidelity due to mode frequency fluctuations with the continuous protocol, \cite{GePRL2019}
\begin{align}
\epsilon_{\text{mf}}=\frac{\pi}{4}\left(\Delta/g\right)^2G^3(1+G).
\label{eq:ard-continuous}
\end{align}
We plot the error for both protocols in Fig. \ref{fig:modeff} for $G=10$. Due to the large factor $C(2r+2)^2$ in Eq. \eqref{eq:ard}, the error from the stroboscopic protocol is much greater than that from the continuous protocol. However, the error due to mode frequency fluctuations in the stroboscopic protocol can be smaller than that of the continuous protocol for very large values of $G$ since the latter grows much faster with $G$ through the dependence of $G^3(1+G)$. By comparing Eqs. \eqref{eq:ard} and \eqref{eq:ard-continuous}, we find this happens for $G>1255$.

For the most realistic conditions ($G\ll1000$), the stroboscopic protocol discussed here requires a more stable motional mode frequency than the continuous protocol. The stroboscopic protocol does have some complementary advantages.  It produces an exponential enhancement, which is important for $gT > 30$.  As shown in Fig. 4, it can also be more effective at reducing the time required to achieve a given geometric phase.  Finally, as detailed in Fig. 4 of the continuous protocol manuscript \cite{GePRL2019}, the enhancements of the continuous protocol can be limited by the breakdown of the RWA.  This is because in the continuous protocol large squeezing requires large detunings between the spin-dependent force and the target mode frequency.  Because the stroboscopic protocol amplifies on resonance with the target mode frequency, larger amplifications are possible without a breakdown of the rotating wave approximation.


\section{Outlook}
In this paper, we proposed a stroboscopic protocol consisting of alternating applications of a spin-dependent force and parametric amplification that amplifies the effective spin-spin interactions in trapped ions induced by the spin-dependent force. Strong parametric amplification can be achieved by modulating the trapping potential at twice the target motional mode frequency. The stroboscopic protocol can lead to an exponential enhancement in the effective coherent interaction strength without requiring stronger laser power in laser-driven trapped-ion gates or larger current in microwave-based trapped-ion gates. Therefore, it can mitigate common challenges, such as the availability of a strong SDF, errors due to spontaneous photon scattering, and couplings to off-resonant, unwanted modes.

Parametric amplification, both through the stroboscopic protocol discussed here or through the continuous protocol discussed previously \cite{GePRL2019}, looks like a useful addition to the ion-trap quantum information processing tool box.  Like most tools in this toolbox, there are trade-offs.  As shown in Secs. III.C and III.D, parametric amplification through either the stroboscopic protocol or the continuous protocol \cite{GePRL2019} can increase errors due to mode frequency fluctuations.  Therefore, when mode frequency fluctuations are the leading source of errors, implementing PA may not make sense.  The impact of other technical errors, such as motional mode heating, need to be considered.
 
As future work, it should be possible to extend the stroboscopic protocol discussed here to enhance the trajectories of multiple modes. For example, by applying several parametric amplifications at different frequencies to selectively amplify different motional modes \cite{Korenblit:2012aa}, respectively. Another interesting direction is to find other useful protocols with more complicated combinations of the SDF and the PA, as sketched in Appendix B.  Protocols employing a phase-insensitive amplification \cite{arenz2018hamiltonian} would also be interesting to investigate, including their sensitivity to mode frequency fluctuations.

\begin{acknowledgments}
We thank David Allcock, Shaun Burd, Murray Holland, Anna Maria Rey,  Athreya Shankar, Daniel Slichter, and James Thompson for interesting discussions. JJB acknowledges financial support from NIST.
\end{acknowledgments}






\appendix
APPENDIX
\section{Parametric Amplification}
In trapped ions, parametric amplification can be done by driving the ion-trap ring-end-cap voltage at close to twice the target mode frequency. Assuming the parametric driving frequency is $\omega_p$, the Hamiltonian describing the $i$th ion motion is \cite{Heinzen1990} $\hat{\mathcal{H}}_{PA}=-\sum_{i=1}^N\frac{2e\hat{z}_i^2}{d^2_T}\text{Re}(Ve^{2i\omega_p t})$, where $V=-i|V|e^{-i\theta}$ and $|V|$ is the voltage and $d_T$ is a characteristic trap dimension. Replacing $z_i$ using the normal-mode expansion, we have
\begin{widetext}
\begin{align}
\hat{\mathcal{H}}_{PA}&=-\frac{2e\text{Re}(Ve^{2i\omega_p t})}{d^2_T}\sum_{i=1}^N\left[\sum_{l,m=1}^Nb_{i,l}b_{i,m}z_{0l}z_{0m}\left(e^{-i\omega_l t}\hat{a}_l+e^{i\omega_l t}\hat{a}_l^{\dagger}\right)\left(e^{-i\omega_m t}\hat{a}_m+e^{i\omega_m t}\hat{a}_m^{\dagger}\right)\right]\nonumber\\
&=-\frac{2e\text{Re}(Ve^{2i\omega_p t})}{d^2_T}\sum_{l,m=1}^N\delta_{lm}z_{0l}z_{0m}\left(e^{-i\omega_l t}\hat{a}_l+e^{i\omega_l t}\hat{a}_l^{\dagger}\right)\left(e^{-i\omega_m t}\hat{a}_m+e^{i\omega_m t}\hat{a}_m^{\dagger}\right)\nonumber\\
&=-\frac{2e\text{Re}(Ve^{2i\omega_p t})}{d^2_T}\sum_{m=1}^Nz_{0m}^2\left(e^{-i\omega_m t}\hat{a}_m+e^{i\omega_m t}\hat{a}_m^{\dagger}\right)^2
\end{align}
\end{widetext}
To obtain the second line, we use the condition $\sum_{i=1}^Nb_{i,m}b_{i,l}=\delta_{ml}$ on the normal-mode transformation matrix. It is interesting to note from the above equation that the Hamiltonian due to PA is independent of the normal-mode transformation on each ion. The parametric driving on the system is just a summation of parametric amplification on each mode separately.  Applying the RWA, we arrive at Eq. \eqref{eq:PA} \begin{eqnarray}
\hat{\mathcal{H}}_{PA}(t)=\sum_{m=1}^N\hat{\mathcal{H}}_{PA}^{(m)}(t),
\end{eqnarray} 
in the main text.

\section{A general protocol}
The most general protocol involving a SDF and PA is to consider alternative applications of the SDF and PA as described below
\begin{align}
\prod_{j=1}^n \hat{\mathcal{U}}_{S}\left(t_{2n+2-2j},t_{2n+1-2j}\right)\hat{\mathcal{U}}_{P}\left(t_{2n+1-2j},t_{2n-2j}\right).
 \label{eq:general}
\end{align}
Note that in each step, the SDF and PA Hamiltonian can be different from previous steps by controlling 
the system parameters, such as $\mu$, $g$, and $\omega_p$. We assume there is only one mode driven by the SDF and PA Hamiltonians. The discussion on the multimode case is given later.

The optimal protocol is to find the maximal geometric area enclosed by the trajectory for fixed parameters $f,\ g, \ T$. However, the optimal solution for a general protocol is difficult to find. To see this, we discuss the constraints for a general protocol involving any combination of a SDF and PA.

We first analyze the optimal protocol in the special case when \emph{only} the SDF is used. The general protocol can be described by applying $n (\ge3)$ displacement $\alpha_k$ in phase space of a motional mode with $\hat{\mathcal{U}}_{S}\left(t_{2n+2-2k},t_{2n+1-2k}\right)=\hat{\mathcal{D}}\left(\alpha_k\right)$. Therefore, we have the following two constraints:
\begin{align}
&\sum_{k=1}^n\alpha_k=0,&\sum_{k=1}^{n}|\alpha_k|\le fT_1,
\end{align}
where the first condition requires the mode undergoes a closed loop in phase space and the second one is a limit on the accumulated path length of a motional mode under a SDF for a time of $T_1$. It can be shown that the optimal protocol using a SDF only is to make a single circle in the phase space according to the isoperimetric inequality. 

For a general off-resonant PA Hamiltonian, the unitary operation can be written as \cite{Carmichael} $\hat{\mathcal{U}}_P\left(t_{2n+1-2j},t_{2n-2j}\right)=\hat{\mathcal{S}}\left(\xi\right)\exp\left(i\varphi\hat{a}^{\dagger}\hat{a}\right)$, where $\xi=re^{i\varphi}$, $\sinh r=\sinh\left(g\tau\sqrt{1-\Delta^2/g^2}\right)/\sqrt{1-\Delta^2/g^2}$, $\tan \varphi=\Delta/\sqrt{g^2-\Delta^2}\tanh\left(g\tau\sqrt{1-\Delta^2/g^2}\right)$, $\tau=t_{2n+1-2j}-t_{2n-2j}$, and $\Delta=\omega_p/2-\omega_1$. Hence, the general protocol using PA and a SDF can be written as
\begin{align}
\label{eq:genseq}
\left[\prod_{k=n}^1\hat{\mathcal{S}}\left(\xi_k\right)\hat{\mathcal{D}}\left(\alpha_k\right)\right]\hat{\mathcal{S}}\left(\xi_0\right),
\end{align}
where we have absorbed the phase rotation $\exp\left(i\varphi\hat{a}^{\dagger}\hat{a}\right)$ into the displacement operations. Rewriting the squeezing operations such that each displacement operator is sandwiched by $\hat{\mathcal{S}}^{\dagger}\left(\xi\right)$ and $\hat{\mathcal{S}}\left(\xi\right)$, we have
\begin{widetext}
\begin{align}
&\left[\prod_{k=n}^0\hat{\mathcal{S}}\left(\xi_k\right)\right]\left[\prod_{k=0}^{n-1}\hat{\mathcal{S}}^{\dagger}\left(\xi_k\right)\hat{\mathcal{D}}\left(\alpha_{n-1}\right)\prod_{k=n-1}^0\hat{\mathcal{S}}\left(\xi_k\right)\right]\cdots\left[\hat{\mathcal{S}}^{\dagger}\left(\xi_0\right)\hat{\mathcal{S}}^{\dagger}\left(\xi_0\right)\hat{\mathcal{D}}\left(\alpha_2\right)\hat{\mathcal{S}}\left(\xi_1\right)\hat{\mathcal{S}}\left(\xi_0\right)\right]
\left[\hat{\mathcal{S}}^{\dagger}\left(\xi_0\right)\hat{\mathcal{D}}\left(\alpha_1\right)\hat{\mathcal{S}}\left(\xi_0\right)\right]\nonumber\\
=&\left[\prod_{k=n}^0\hat{\mathcal{S}}\left(\xi_k\right)\right]\prod_{k=n}^1\hat{\mathcal{D}}\left(\alpha_kd_k\right),
\end{align}
\end{widetext}
where the second line is obtained using the relation $\hat{\mathcal{S}}^{\dagger}\left(\xi\right) \hat{a}\hat{\mathcal{S}}\left(\xi\right)=\cosh r \hat{a}-\sinh r e^{i\varphi}\hat{a}^{\dagger}$ \cite{SZ}. Therefore, each displacement operation $\alpha_k$ is amplified and rotated with a magnitude of $|d_k|(\ge1)$ and an angle $\text{arg}[d_k]$, respectively. After the PA and SDF operations, we require that the mechanical state returns to its initial state. In this case, we have three constraints:
\begin{align}
&\sum_{k=1}^n\alpha_k d_k=0,&\prod_{k=1}^{n}|d_k|\le \exp(gT_2), \ \ &\sum_{k=1}^{n}|\alpha_k|\le fT_1,\label{eq:constraints}
\end{align}
where the first equation is satisfied for closing the loop under the SDF and PA, and the second constraint is such that the accumulated squeezing is bounded by the total squeezing if accumulated in a given amount of time $T_2$. The gate time is $T=T_1+T_2$. Additionally, we would require a condition on the overall squeezing given by the PA unitary operations to be zero such that the motional state is back to an unsqueezed state at the gate, i.e., $\prod_{k=n}^0\hat{\mathcal{S}}\left(\xi_k\right)=1$.
Our goal is to optimize the area enclosed by the sides $\alpha_kd_k$ under the constraints in Eq. \eqref{eq:constraints} for a given set of parameters $(f,g,T)$. All together we will have at least $4n-4$ variables, assuming the inequalities become equalities. The optimal solution for fixed $n$ is not easy to find since the minimum number of variables is $8$. The optimal solution for an arbitrary $n$ is therefore more difficult.

\section{Error Analysis}
\subsection{Off-resonant analysis}
Now we discuss the situation when the target mode is off resonant to the SDF and the PA in the stroboscopic protocol while all the other parameters are ideal. This is due to the motional mode frequency fluctuation. As an example, we analyze specifically the square protocol. The sequence of operations from right to left and from top to bottom is given by
\begin{widetext}
\begin{align}
&\hat{\mathcal{U}}_P\left(2t_1+4t_2,2t_1+3t_2\right)\hat{\mathcal{U}}_{S}\left(2t_1+3t_2,t_1+3t_2\right)\hat{\mathcal{U}}_P^{\dagger}\left(t_1+3t_2,t_1+2t_2\right)\hat{\mathcal{U}}_P^{\dagger}\left(t_1+2t_2,t_1+t_2\right)\hat{\mathcal{U}}_{S}\left(t_1+t_2,t_2\right)\hat{\mathcal{U}}_P\left(t_2,0\right)\nonumber\\
&\hat{\mathcal{U}}_P^{\dagger}\left(3t_1+6t_2,3t_1+5t_2\right)\hat{\mathcal{U}}_{S}\left(3t_1+5t_2,2t_1+5t_2\right)\hat{\mathcal{U}}_P\left(2t_1+5t_2,2t_1+4t_2\right)\times\nonumber\\
&\hat{\mathcal{U}}_P\left(4t_1+8t_2,4t_1+7t_2\right)\hat{\mathcal{U}}_{S}\left(4t_1+7t_2,3t_1+7t_2\right)\hat{\mathcal{U}}_P^{\dagger}\left(3t_1+7t_2,3t_1+6t_2\right)\times
\label{eq:non-resonant}
\end{align}
\end{widetext}
where $t_1$ and $t_2$ are the interaction times for one operation of the SDF and the PA, respectively. The non-zero detuning of the target mode result in three effects: i) rotation of the displacement applied by the SDF; ii) rotation of the squeezing angle and iii) reduction of the squeezing strength by the PA. For a small detuning $\Delta=\mu-\omega_1\ll g$, these effects are small by themselves, i.e., i) and ii) depend linearly on $\Delta$, and iii) has a quadratic relation with $\Delta$. However, a pair of squeezing and anti-squeezing operations no longer cancel each other exactly, for example $\hat{\mathcal{U}}_P\left(2t_1+4t_2,2t_1+3t_2\right)$ and $\hat{\mathcal{U}}_P^{\dagger}\left(t_1+3t_2,t_1+2t_2\right)$, which can lead to an exponential amplification on the error. To illustrate this idea, we consider a pair of squeezing and anti-squeezing operations after a displacement operation given by
\begin{align}
\hat{\mathcal{S}}\left(re^{i\epsilon_2}\right)\hat{\mathcal{S}}\left(-re^{i\epsilon_1}\right)\hat{\mathcal{D}}\left(\alpha\right),
\end{align}
where we neglect the phase term from the non-resonant PA operation by considering just the squeezing operation (see Eq. \eqref{eq:genseq}). Here $\epsilon_1,\epsilon_2\sim \Delta/g\ll1$ are the rotation of the squeezing axis due to the detuning. Without loss of generality, we assume $\alpha$ to be real and obtain
\begin{align}
\hat{\mathcal{S}}\left(re^{i\epsilon_2}\right)\hat{\mathcal{S}}\left(-re^{i\epsilon_1}\right)\hat{\mathcal{D}}\left(\alpha\right)\approx \hat{\mathcal{D}}\left[\alpha\left(1+i\frac{\epsilon_1-\epsilon_2}{2}e^{2r}\right)\right].
\label{eq:ff-amp}
\end{align}
Due to the time dependence on the rotation, $\epsilon_1\ne\epsilon_2$ in general, therefore the residual displacement is on the order of $(\Delta/g )e^{2r}$. The corresponding error in the gate fidelity is reduced by a number on the order of $(\Delta/g )^2e^{4r}$. However, we note that the geometric phase error is negligible comparing to the residual displacement error for $\Delta/g\ll1$ from our numerical simulations.

The above analysis can be adapted to discuss the residual displacement of a spectator mode that is far-off resonant, i.e. $\Delta_m\gtrsim g$. In this case, the squeezing strength on the spectator mode is greatly suppressed and the residual displacement can be made small since we can use a smaller SDF $f$ enabled by the parametric amplification. We plot in Fig. \ref{fig:ratio}  the ratio of displacement with and without the PA after one-leg of displacement amplification $\hat{\mathcal{U}}_P^{\dagger}\left(t_1+2t_2,t_1+t_2\right)\hat{\mathcal{U}}_{S}\left(t_1+t_2,t_2\right)\hat{\mathcal{U}}_P\left(t_2,0\right)$ numerically for different values of $\Delta_m$.

\subsection{Phase fluctuations}
For a phase uncertainty between the PA and the SDF, the analysis is similar to that of the frequency fluctuation. Assume phase fluctuations $\Delta\theta$ in one operation of the PA, they may lead to a displacement error amplified by $e^{2r}$ according to Eq. \eqref{eq:ff-amp} as
\begin{align}
\hat{\mathcal{S}}\left(re^{i\Delta\theta}\right)\hat{\mathcal{S}}\left(-r\right)\hat{\mathcal{D}}\left(\alpha\right)\approx \hat{\mathcal{D}}\left[\alpha\left(1+i\frac{\Delta\theta}{2}e^{2r}\right)\right].
\label{eq:pf-amp}
\end{align}
Hence we also require very stable phase alignment between the PA and the SDF.

\subsection{Timing control fluctuations}
We consider the situation when the fluctuations happen in the one side of the operations for the SDF or the PA. In the operation of the SDF, the errors are larger when fluctuations occur in the second and the third sides of the operation than in the other sides. Considering the second side for example, the displacement operator becomes $\hat{\mathcal{D}}\left[i\alpha e^r(1+\epsilon)\right]$, where $\epsilon\equiv \Delta t/t_1$. Therefore, $\Delta \alpha\sim \epsilon\sqrt{\Phi}$ and $\Delta\Phi\sim\epsilon \Phi$. In the operation of the PA, timing fluctuations affect the amplification on the displacement, for example $\hat{\mathcal{D}}(\alpha e^r)$ is replaced by $\hat{\mathcal{D}}(\alpha e^{r(1+\epsilon)})$. So there is only uncertainty in the residual displacement, i.e., $\Delta \alpha\sim \left(e^{r\epsilon}-1\right)\sqrt{\Phi}\sim r\epsilon\sqrt{\Phi}$, which can be amplified from the PA. Moreover, the time fluctuation in PA may lead to a squeezed-displaced final state $\ket{\epsilon,\alpha\epsilon}_c=\hat{\mathcal{D}}\left(\alpha e^r r\epsilon\right)\hat{\mathcal{S}}\left(r\epsilon\right)\ket{0}_c$ that is entangled with the spin states. This can lead to a reduced fidelity of a target state since $\left|\braket{0|\epsilon,\alpha\epsilon}_c\right|^2\approx 1- \left(\alpha e^{r}r \epsilon\right)^2(1-r \epsilon)$ \cite{SZ}.

\section{Lamb-Dicke limit}
Due to the mechanical squeezing, the condition on the Lamb-Dicke regime will be revised, which can lead to limitations on how strongly we can amplify. As can be seen in the actual trajectory in Fig. \ref{fig:sp} (a), the maximum displacement is on the order of $\alpha e^{2r} \sqrt{\braket{\hat{S}^2_z}}$, where $\hat{S}_z\equiv1/2\sum_i\hat{\sigma}_i^z$. With the Lamb-Dicke limit, we have the following requirement on the parametric amplification for the square protocol \cite{GePRL2019}
\begin{align}
e^{r}\ll \frac{1}{\eta}\sqrt{\frac{N}{2\Phi \braket{\hat{S}^2_z}}},
\end{align}
where the optimal $r=gT/4-1$ is assumed. The maximum value of $gT$ permitted by Lamb-Dicke confinement depends on the experiment and the type of spin states we would like to create.


\bibliography{Improving-Trapped-Ion-quantum-simulators-via-Parametric-Amplification}

\end{document}